\documentstyle[12pt]{article} 
\def\be{\begin{equation}}
\def\ee{\end{equation}}
\def\ba{\begin{eqnarray}}
\def\ea{\end{eqnarray}}

\def\negenspace{\kern-1.1em}
\def\quer{\negenspace\nearrow}

\begin{document}



\title{Chiral supergravity and anomalies} 
 
\author{Eckehard W. Mielke\thanks{E-mail: ekke@xanum.uam.mx} $\>$  and
Alfredo Mac\'{\i}as\thanks{E-mail: amac@xanum.uam.mx}\\ 
Departamento de F\'{\i}sica,\\ 
Universidad Aut\'onoma Metropolitana--Iztapalapa,\\ 
P.O. Box 55-534, 09340 M\'exico D.F., MEXICO.} 
 
\date{\today} 

\maketitle 

\begin{abstract}

Similarly as in the Ashtekar approach, 
the {\em translational Chern--Simons term} is,  as
a {\em generating function}, instrumental for 
a {\em chiral reformulation} of  {\em simple supergravity}. 
After applying the algebraic Cartan relation between spin and torsion, 
 the resulting canonical transformation induces not only  
decomposition  of the gravitational fields into selfdual and antiselfdual 
modes, but also a splitting
of the Rarita--Schwinger  fields into their chiral parts 
in a natural way. In some detail, we also analyze the consequences for 
{\em axial} and {\em chiral} anomalies. 

\end{abstract}

\vspace{.2cm} 

\begin{flushleft} 
{PACS numbers: 04.60.Ds, 04.20.Fy, 04.50.+h} 

\bigskip

KEYWORDS: Supergravity, Duality, Anomalies
\end{flushleft} 

\section{Introduction} 
 
Recent developments by Ashtekar \cite{ash,ASH}, constitute 
considerable progress in the canonical formulation of general relativity and 
its relation to  $SU(2)$ Yang--Mills theory. The key feature of Ashtekar's 
formulation of general relativity is the introduction of a {\em selfdual} 
connection as one of the basic dynamical variables. This allows us to 
formulate general relativity in the familiar phase space of the Yang--Mills
theories. It is indeed a theory of a connection and proves to be very useful 
to have the possibility of using loop variables (Wilson loops of the Ashtekar 
selfdual connection) \cite{rovsmo} in both the classical and the quantum 
descriptions of general relativity. This new formulation has generated  
not only a new approach to the problem of formulating quantum theory 
\cite{rov}, but opened up a new avenue to 
study quantum gravity from a nonperturbative point of view. Since  a great 
deal of nonperturbative effects in the Yang--Mills theories is known, 
it seems now feasible
to carry over several techniques from quantum Yang--Mills theories
to quantum gravity.

On the other hand, the operation of 
decomposing a field into positive and negative frequencies
has {\em no} counterpart beyond linearized theory,
while  decomposing a field into selfdual and antiselfdual pieces  
makes sense for both {\em non--linear} gauge fields and general relativity.  
The role of {\em chirality} can be identified with the concept of 
{\em selfduality}. 
This is how, concretely, selfduality might be significant for quantizing
either gauge theories or gravity.

In quantum theory one commonly {\em chooses} to work
with positive frequency fields: positive helicity photons
correspond to selfdual fields, whereas negative helicity ones
to anti--selfdual fields.
Alternatively, however, one could adopt {\em only} selfdual fields. 
The positive frequency fields would yield in this case the helicity 
$+1$ photons and the negative frequency fields the helicity $-1$  
photons. 

In spite of all the success of the Ashtekar formulation, the theory gets 
afflicted by the complex character of the variables, thus the issue of the
reality conditions becomes compulsory. 
In the classical theory, general relativity is embbeded in a larger complex
theory \cite{karel}.

As it is well known, supergravity suffers the same diseases as general 
relativity. Had we aimed at giving to the supergravity Lagrangian its chiral 
form, a similar analysis to the one given by Ashtekar should be performed.  
Jacobson \cite{Jacob88} advanced to some extent in this direction. Here we 
will turn, however, to the Clifford--algebra approach which yields the chiral  
decomposition from the more fundamental point of view of the 
{\em generating function} \cite{ma96}.

In this paper, we introduce a purely imaginary {\em translational}  
Chern--Simons  term as a generating function which induces the {\em chiral} 
decomposition into selfdual and anti--selfdual parts in the bosonic sector as 
well as in the fermionic sector of {\em simple}, i.e. (${\cal N}$=1) supergravity.

The chiral form of Einstein--Cartan theory of gravity and its coupling 
to Rarita--Schwinger field turn out to be very simple when expressed in the 
language of Clifford algebras. For instance, the coupling necessary to 
implement the local supersymmetry of the (${\cal N}$=1) supergravity action is a 
natural step in the Clifford formulation of the theory.

The crucial point is here to add suitable boundary terms of the
{\em Chern--Simons} type to the action which allow for a {\em chiral}  
formulation; this induces in a natural way the introduction of an
anti--selfdual or selfdual connection.
In the Hamiltonian formulation such a boundary term acts as 
{\em generating function} for new variables, see \cite{[13],ma96}.
The structure of these boundary terms is fixed by the requirement  
of local Lorentz invariance and their relation to the two Bianchi identities. 

The plan of the paper is as follows: In Sec. 2 we present the  necessary  
requisites of the geometrical setting in terms of 
Cliffords algebra--valued forms.  The
generating 
function of the Chern--Simons type boundary terms is discussed in Sec. 3.
In Sec. 4, we revisit the simple  supergravity.
In Sec. 5,  {\em chiral}  supergravity is generated by the `on shell' 
boundary term.
In Sec. 6, the chiral anomaly is investigated and some controverses are resolved.
In Sec. 7 we explain the issue of the reality constraint in terms of
 a toy model and finally 
in Sec. 8 our  results are contrasted with other approaches.

\section{Riemann--Cartan spacetime and  Clifford algebras}
The Riemann--Cartan (RC) geometry \cite{Hehl76} will be described by means of a very  
concise formalism employing {\em Clifford algebra--valued 
exterior differential  
forms}. 
The Dirac matrices $\gamma_{\alpha}$ obey
\begin{equation} 
\gamma_\alpha\gamma_\beta+\gamma_\beta\gamma_\alpha 
  =2o_{\alpha\beta} \, {\bf 1}_4
\label{gamma5}\,, 
\end{equation} 
where $\alpha, \beta,\cdots$=\^0, \^1, \^2, \^3 denote the (anholonomic) 
indices of the frame field $e_\alpha$ which is assumed to be 
{\em orthonormal}. 
The signature of the Minkowski metric $o_{\alpha\beta}$ of the frame 
bundle is $(+---)$. 
The 16 matrices $\{{\bf 1}_4,\gamma_\alpha,\sigma_{\alpha\beta},\gamma_5, 
\gamma_5\gamma_\alpha\}$ form a basis of a {\em Clifford algebra} in 
four dimensions, see \cite{[16],Bay}. 
The totally antisymmetric product of Dirac matrices  
is the zero--form 
$\gamma_5:= (i/4!)\;^*(\gamma\wedge\gamma\wedge\gamma\wedge\gamma)$ and 
fulfills 
\be 
\gamma_5=-i\,\gamma_{\hat{0}}\,\gamma_{\hat{1}}\,\gamma_{\hat{2}}\,
\gamma_{\hat{3}}\,,\quad\gamma_5\, \gamma_5=+{\bf 1}_4\,, \quad
\{\gamma_5,\,\gamma_\alpha\}=0
\label{ecardo}\, . 
\ee 

With respect to the {\em trace} $Tr$, the elements of the Clifford algebra  
are normalized by 
\begin{equation} 
  Tr(\gamma_\alpha\; \gamma_\beta) =4\,o_{\alpha\beta} \qquad {\rm and} 
  \qquad Tr(\sigma_{\alpha\beta}\; \sigma^{\gamma\delta}) = 
  8\,\delta_{[\alpha}^{\gamma}\; \delta_{\beta]}^{\delta} 
\label{trace} \, , 
\end{equation} 
where $[\alpha\, \beta ]=\frac{1}{2}(\alpha\beta-\beta\alpha)$ denotes the 
antisymmetrization of indices. 
Following the notation of Ref. \cite{[17],[18],[15]}, the constant
$\gamma_\alpha$ matrices can be converted into Clifford
algebra--valued coframe one-- or three--forms, respectively: 
\begin{equation} 
\gamma:=\gamma_\alpha\vartheta^\alpha\,, 
 \qquad\;{^{*}\gamma}=\gamma^\alpha\, 
  \eta_\alpha = {i\over 6}\,{\gamma_5}\;\gamma\wedge\gamma\wedge\gamma\,. 
\label{cofr} 
\end{equation} 

Here, we restrict ourselves to a to\-polo\-gically trivial frame bundle where 
$\eta= (1/4!)\eta_{\alpha\beta\gamma\delta}\; \vartheta^\alpha\wedge 
 \vartheta^\beta\wedge  \vartheta^\gamma\wedge \vartheta^\delta$ is  
the volume four--form with the normalization 
$\eta_{\hat{0}\hat{1}\hat{2}\hat{3}}= +1$,  
$\eta_\alpha:=e_\alpha\rfloor\eta= \;^{*}\vartheta_\alpha$ is the 
coframe ``density'', and $^*$ the Hodge dual. Moreover, in order to 
complete the $\eta$--basis for forms, we define $\eta_{\alpha\beta} 
:=e_\beta \rfloor\eta_\alpha $, $\eta_{\alpha\beta\gamma} :=e_\gamma 
\rfloor \eta_{\alpha\beta}$, $\eta_{\alpha\beta\gamma\delta} 
:=e_\delta \rfloor \eta_{\alpha\beta\gamma} $. 
The contraction operator acting on p\--forms from the left is defined as 
\begin{equation} 
\check{\gamma}:=\gamma^{\beta}e_{\beta}= 
{\gamma}^{\beta}e^{i}{}_{\beta}\,{\partial}_{i} \qquad {\rm with}  
\qquad 
\check\gamma\rfloor \gamma =4\cdot {\bf 1}_4 \, . 
\label{contract} 
\end{equation} 
It generalizes the usual Feynman ``dagger" convention $A\quer := 
\gamma^\alpha e_\alpha\rfloor A$ for one--forms $A$.

The Lorentz generators $\sigma_{\alpha\beta}:={i\over 2} 
(\gamma_\alpha\gamma_\beta-\gamma_\beta\gamma_\alpha)$ obey 
\begin{equation} 
  [\gamma_\alpha,\sigma_{\beta\gamma}]= 
2i\,(o_{\alpha\beta}\,\gamma_\gamma-o_{\alpha\gamma}\,\gamma_\beta)\,, 
 \qquad 
\{\gamma_\alpha\,, \sigma_{\beta\gamma}\} 
  =2i\,\gamma_{[\alpha}\,\gamma_\beta\,\gamma_{\gamma]}= 
-2\eta_{\alpha\beta\gamma\delta} \gamma_{5}\gamma^{\delta} \, . 
\label{anticomm} 
\end{equation} 
Their {\em Lie} (or right) duals and their inverses are related via 
\begin{equation}
\sigma^{\star}_{\alpha\beta}:= 
{1\over 2}\,\sigma^{\gamma\delta}\,\eta_{\alpha\beta\gamma\delta}=
  i\gamma_5 \,\sigma_{\alpha\beta}, \qquad 
 {\sigma}_{\alpha\beta}=-{1\over 2}{\sigma}^{\star}_{\mu\nu} 
    {\eta}^{\mu\nu}\, _{\alpha\beta}
\label{liedual}\, . 
\end{equation} 
The associated two-forms are given by 
\begin{equation} 
\sigma:={1\over 2}\sigma_{\alpha\beta}\,\vartheta^\alpha\wedge 
  \vartheta^\beta={i\over 2}\,\gamma\wedge\gamma\,, \qquad 
  \,^*\sigma={1\over 2}\sigma_{\alpha\beta}\,\eta^{\alpha\beta} 
  =:\sigma^{\star}= i\,\gamma_5\,\sigma 
\label{sig}\, . 
\end{equation} 

Note that in orthonormal frames the Hodge dual $^*$ and the Lie dual $^\star$ 
are identical operations for $\sigma$. For Minkowski signature, the Hodge dual
satisfies $^{**}=-1$, therefore often $i\,^*$ is used in field theory 
in order to have an
{\em involutive} duality operator.
We will encounter also the self-- or antiself dual  combination 
\be\sigma_\pm:= (\sigma \pm i \,^*\sigma)/2 = {1\over 2}(1\mp \gamma_5)\, 
\sigma,\quad {\rm with} \quad i \,^*\sigma_\pm = \pm \sigma_\pm
\ee
which at times is referred to as the {\em Pleba\v nski two--form} \cite{Pleb}, 
cf. also Debever \cite{de64} and Brans \cite{bra} for related 
earlier constructions. 
\subsection{Clifford algebra-valued torsion and curvature}
In terms of the Clifford algebra--valued {\em connection} 
$\Gamma := {i\over 4} \Gamma^{\alpha\beta}\,\sigma_{\alpha\beta}$, the 
$\overline{SO}(1,3)\cong SL(2,C)$--covariant 
exterior derivative $D=d+[\Gamma,\quad]$ employs the algebra--valued 
{\em form commutator} $[\Psi, \Phi] := 
\Psi\wedge\Phi - (-1)^{pq} \Phi\wedge\Psi$. Differentiation of the basic 
variables leads to the Clifford
algebra--valued {\em torsion} and {\em curvature} two--forms, respectively: 
\begin{equation} 
\Theta :=D\gamma =T^{\alpha}\gamma_{\alpha}\;, \qquad 
\Omega := d\Gamma +\Gamma\wedge \Gamma = 
{i\over 4}R^{\alpha\beta}\,\sigma_{\alpha\beta}\, . 
\label{tor} 
\end{equation} 
In accordance with (\ref{liedual}), its Lie dual is given by 
\be
\Omega^\star:=\frac{i}{8} R_{\alpha\beta}\,  \eta^{\alpha\beta\gamma\delta}
\sigma_{\gamma\delta} = - \frac{1}{4} R^{\alpha\beta} \gamma_5 
\sigma_{\alpha\beta}= i\gamma_5 \Omega \, .
\ee
The Ricci identity reads 
\be 
D D \Psi = [\Omega,\Psi] 
\label{ric} \, . 
\ee 

In this ``Clifform'' approach, the torsion two--form can be irreducibly  
decomposed into the trace part 
$\,^{(2)}\Theta$, the axial torsion $\,^{(3)}\Theta$, and the 
tensor torsion  $\,^{(1)}\Theta$ as follows:
\ba \,^{(2)}{\Theta} & := & 
{1\over 3}\,\gamma\wedge T 
= {1\over 12}\,\gamma\wedge \, ^* Tr(\, ^*\Theta\wedge \gamma)\,,
\label{trator}\\ 
\,^{(3)}{\Theta} & := & 
-{1\over 3}\,
{}^*\left[\gamma\wedge A \right] 
= {1\over 12}\,\check{\gamma}\rfloor Tr(\gamma\wedge\Theta)\,,\\ 
\,^{(1)}{\Theta} & := & \Theta -\,^{(2)}{\Theta}-\,^{(3)}{\Theta} 
\label{vector}\,. 
\ea 
The one--forms of the trace and axial vector torsion, respectively, are 
defined by
\be T:= {1\over 4}\, Tr\left(\check{\gamma} \rfloor\Theta\right) =e_\alpha\rfloor T^\alpha\,, 
\qquad  A:=
{1\over 4}\,^*Tr(\gamma\wedge\Theta) =   
\,^*(\vartheta_\alpha\wedge T^\alpha)\, .
\ee

The purely geometrical identity \cite{HMMN}
\begin{equation} 
 {\Theta}\wedge {\Theta} \equiv \,^{(1)}\Theta\wedge  \,^{(1)}\Theta 
 + 2 \,^{(2)}\Theta\wedge  \,^{(3)}\Theta \, , 
\label{tortor} 
\end{equation} 
vanishes for purely vector or axial torsion. 
%

\section{Chern--Simons type boundary terms in gravity}
  
Quite generally,  the Chern--Simons  
term 
\be 
C := Tr\, \left(A\wedge F -{1\over 3} A\wedge A \wedge A\right)\, , 
\label{2.26} 
\ee 
of a non--Abelian gauge theory  yield  boundary terms $dC$ which 
turn out to be {\em generating functions}.
In the Hamiltonian formulation, they induce a new pair of canonical variables, 
see for instance Mielke \cite{[27],[13]} in the case of the  
Ashtekar reformulation of GR. 

In a Riemann--Cartan geometry, the {\em Chern--Simons term} for the  
Lorentz connection reads 
\begin{equation} 
C_{\rm RR}  := - Tr\, \left( {\Gamma}\wedge {\Omega} - 
{1\over 3} {\Gamma}\wedge {\Gamma}\wedge  {\Gamma} \right)  \, , 
\label{CRR} 
\end{equation} 
whereas the {\em translational} Chern--Simons term 
can be written as 
\begin{equation} 
C_{\rm TT}  :=  {1\over{8\ell^2}} Tr\, ( {\gamma} \wedge  {\Theta} )= 
{1\over{2\ell^2}}\, {\vartheta^\alpha}\wedge T_{\alpha}\,.  
\label{CTT} 
\end{equation} 
A fundamental length $\ell$ introduced here is necessary for 
dimensional reasons.

The corresponding boundary terms $dC$ are distinguished by the  
fact that their variational derivatives 
\be 
{ {\delta }\over{\delta\gamma} } dC_{\rm TT}= 
{1\over\ell^2}\,\left( D\Theta- [\Omega, 
\gamma]\right)\qquad {\rm and} \qquad 
{ {\delta}\over{\delta\Gamma} } dC_{\rm RR} = - D\Omega \, ,
\ee 
lead us back to the first and second Bianchi identity
\begin{equation} 
D\Theta \equiv [\Omega ,\gamma]\, ,\qquad\qquad 
D\Omega\equiv 0 
\label{Bia}\, , 
\end{equation}  
respectively \cite{HKMM}. 

The contracted Bianchi identities 
\begin{equation} 
D[\gamma\, ,\Theta ]\equiv 2i[\sigma\,,\Omega]\, ,\qquad 
D[\gamma\, ,\Omega ]\equiv [\Theta\,, \Omega]\, ,\qquad  D{\cal E}\equiv 
i[\Theta,\gamma_{5}\Omega] 
\label{CBia}\, , 
\end{equation}
ensure the automatic conservation of   the {\em Einstein three--form}
\begin{equation} 
{\cal E}:= {\cal E} _{\alpha}{\gamma}^{\alpha} := {1\over 2}R^{\mu\nu}\wedge 
{\eta}_{\mu\nu\lambda} 
{\gamma}^{\lambda} = -i\,{\gamma}_{5}( {\Omega}\wedge {\gamma} + 
{\gamma}\wedge{\Omega})=i[\gamma,\gamma_{5}\Omega] 
\label{E3} 
\end{equation}
in Einstein's GR with $\Theta=0$.  
\subsection{Chiral decomposition of the connection}

The translational Chern--Simons term {\em induces} a chiral  
representation of the supergravity (Riemann--Cartan) curvature and, in the 
wake of this, two chiral pieces of the connection are projected out. The 
self-- or antisel dual part of the Lorentz  connection is defined according to
\be
\Gamma={\Gamma}_{+} + {\Gamma}_{-} \,, \qquad
\Gamma_\pm^{\alpha\beta}:={1\over
  2}\,\Gamma^{\alpha\beta} \pm {i\over 2}
\eta^{\alpha\beta\gamma\delta}\Gamma_{\gamma\delta}\,.
\ee
This decomposition of the connection results also from the action of the 
 corresponding projection operator $P_\pm={1\over 2}(1\pm \gamma_5)$, the  
chirality operator:
\begin{equation}
\Gamma_\pm=
P_\mp\Gamma = {i\over 8} \Gamma^{\alpha\beta}(1\mp\gamma_5)
\sigma_{\alpha\beta}=
{i\over 8} \Gamma^{\alpha\beta}(\sigma_{\alpha\beta} \pm
 i\,\sigma^{\star}_{\alpha\beta})  = {i\over 4}  
\sigma_{\alpha\beta}\left(
{1\over 2}\Gamma^{\alpha\beta} \pm {i\over 2}
\eta^{\alpha\beta\gamma\delta}\Gamma_{\gamma\delta}\right)\,.
\end{equation}
 whereas the basis one--form partially anticommutes with the chiral projection
\begin{equation}
\gamma_\pm:=
P_\mp\gamma = {1\over 2}(1\mp \gamma_5)\gamma = 
{1\over 2}\gamma(1\pm \gamma_5)= \gamma P_\pm\, .
\end{equation}
 Note that the
chirality operator $P_{\pm}$ of our Clifford approach is the same as 
the one used in  elementary particle physics.

Observe also that ${\Gamma}_-$ has tensorial transformation
properties.
Substituting the connection into the curvature, we find a corresponding 
decomposition of the curvature:
\begin{equation}
  \Omega=\Omega_++\Omega_-\,,\qquad
  \Omega_\pm:=\Omega(\Gamma_\pm)\,.\label{curvdual}
\end{equation}
into selfdual and antiselfdual pieces \cite{MMM96}.

The connection dynamics looks in many respects simpler than usual
geometrodynamics based on the metric, but the simplicity has been bought at a price that the
connection $\Gamma_{\pm}$ is necessarily complex.

\section{Simple (${\cal N}$=1) supergravity} 

{\em Supergravity}  \cite{New81,Freed} 
with one supersymmetry generator, i.e. ${\cal N}$=1, represents 
the simplest consistent coupling of a Rarita--Schwinger (RS) type 
spin--$\frac{3}{2}$ field to gravity.  

The corresponding Hermitian Lagrangian four--form reads 
\begin{eqnarray} 
 L_{\rm Sugra} &=&V_{\rm EC} + L_{\rm RS} \label{eq:sugra} \, , \\ 
 L_{\rm RS}  &=& -\frac{1}{2} \left({\overline{\Psi}}\wedge 
                   \gamma_5\gamma\wedge D\Psi-\overline{D\Psi}\wedge 
                   \gamma_5\gamma\wedge\Psi\right)\, , 
\label{eq:rsch} 
\end{eqnarray} 
where 
\be 
V_{\rm EC} = {i\over{2\ell^2}}\, Tr\, (\Omega\wedge\,^* \sigma ) = - 
{1\over{2\ell^2}}\, Tr\, (\Omega\wedge \gamma_5\,\sigma )\,.
\label{ECL} 
\ee 
denotes the Einstein\---Cartan (EC) Lagrangian.

It should be noted that the Rarita--Schwinger field 
$\Psi:=\Psi_{\alpha}\vartheta^{\alpha}$ entering Eq. (\ref{eq:rsch})  
is a {\em Majorana spinor} valued one-form. 
As it is well known, it satisfies the {\rm Majorana condition}, 
i.e. $\Psi=C\overline{\Psi}{}^{t}$, where 
$\overline{\Psi}:=\Psi^\dagger\,C$, and the charge conjugation matrix 
\cite{ma} given by $C=-i\gamma^{\hat 0}$ satisfies 
$C^\dagger=C^{-1}\,,\quad C^{t}=-C\,,\quad C^{-1}\gamma^\alpha C 
=-\left(\gamma^\alpha\right)^{t}$. 
Consequently, 
\be 
\overline{\Psi}\wedge\Psi=0\,,\quad\overline{\Psi}\wedge\gamma_5 
\gamma^\alpha\Psi=0\,, \quad 
\overline{\Psi}\wedge\gamma_5\Psi=0\, . 
\label{cons}
\ee 
Here we will use the real Majorana representation in  
which all 
$\gamma^{\alpha}$ are purely imaginary but the components of the  
gravitino 
vector--spinor are consequently all real, see \cite{kak,ma}. 

The covariant exterior derivatives  
\be 
D\Psi:=d\Psi+\Gamma\wedge\Psi\,,\qquad 
\overline{D\Psi}=d\overline{\Psi}+\overline{\Psi}\wedge \Gamma 
\label{covex}\, 
\ee
of spinor--valued one--forms, induce minimal coupling, contrary to the tensorial
component notation \cite{ramond}. 

Energy--momentum and spin currents of the Rarita\---Schwinger field turn out 
to be 
\be 
\Sigma_\alpha={1\over 2}\,\left(\overline{\Psi}\wedge\gamma_5 
\gamma_\alpha D\Psi+\overline{D\Psi}\wedge\gamma_5\gamma_\alpha\Psi\right) 
\label{em1}\, 
\ee 
and 
\begin{equation} 
\tau_{\alpha\beta}:= 
\frac{\partial L_{\rm RS}}{\partial \Gamma^{\alpha\beta}}= 
- {i\over 8}\overline{\Psi}\wedge\gamma_5 
\left(\gamma\sigma_{\alpha\beta}+\sigma_{\alpha\beta}\gamma\right) 
\wedge\Psi={i\over 4}\,\eta_{\alpha\beta\gamma}\wedge 
\overline{\Psi}\wedge\gamma^\gamma\Psi\, , 
\label{sp} 
\end{equation} 
respectively. 

With our Clifford definition (\ref{tor}) for the torsion, we find for  
the {\em Rarita--Schwinger  equation}, cf. \cite{Ur},
\be 
\gamma\wedge D\Psi-{1\over 2}\,\Theta\wedge\Psi=0\, . 
\label{RS-EC'} 
\ee 
For the coupled Einstein--Cartan--Rarita--Schwinger Lagrangian, 
the first field equation reads 
\be 
i\,{\gamma}_{5}( {\Omega}\wedge {\gamma} + {\gamma}\wedge{\Omega}) 
= {\ell^{2}\over 2}\,\gamma^\alpha \left(\overline{\Psi}\wedge\gamma_5  
\gamma_\alpha D\Psi+\overline{D\Psi}\wedge\gamma_5\gamma_\alpha\Psi\right)
\label{fir'}\,, 
\ee 
whereas the second field equation, 
our ``Cartan relation", turns out to be: 
\be 
-{1\over 2}\,\eta_{\alpha\beta\gamma}\wedge T^\gamma={i\over 4}  
\ell^{2}\, 
\eta_{\alpha\beta\gamma}\wedge \overline{\Psi}\wedge\gamma^\gamma\Psi 
\label{sec1}\, ,
\ee 
or, more simply written, 
\be 
\Theta=T^{\alpha}\gamma_{\alpha}= -{i\over 2}\ell^{2}\, 
\overline{\Psi}\wedge\gamma^{\alpha}\Psi\gamma_{\alpha} 
\label{sec2}\,. 
\ee 
For the explicit torsion term in (\ref{RS-EC'}), we find 
\be 
\Theta\wedge\Psi=-\frac{i}{2}  
{\ell}^{2}\overline{\Psi}\wedge\gamma^{\alpha} 
\Psi\wedge\gamma_{\alpha}\Psi =0 
\label{torcon}\, , 
\ee 
as can be easily shown by means of a commutation of the one--form 
$\gamma^{\alpha}\Psi$ among itself, i.e.
a Fierz reordering \cite{New81}. 
Thus, in sharp contrast to the Dirac case, for the supergravity coupling of 
the Rarita--Schwinger field, torsion does {\em not} induce a 
nonlinear term in (\ref{RS-EC'}). This is because the RS equation amounts to 
four Dirac equation plus the constraint (\ref{torcon}).
Covariant exterior differentiation of the Rarita--Schwinger equation 
(\ref{RS-EC'}) yields: 
\ba 
D\left(\gamma\wedge D\Psi-{1\over 2}\,\Theta\wedge\Psi\right) 
&=&\Theta\wedge D\Psi - \gamma\wedge DD\Psi - {1\over 2}[\Omega,\gamma]\wedge 
\Psi -{1\over 2}\Theta\wedge D\Psi \nonumber \\ 
&=&{1\over 2}\,\Theta\wedge D\Psi - {1\over 2}\,\left( 
{\Omega}\wedge {\gamma} + {\gamma}\wedge{\Omega}\right)\wedge \Psi 
\label{DRS''''}\,. 
\ea 
It is a remarkable fact of supergravity also in higher dimensions, 
cf. \cite{BTZ96}, that the {\em integrability condition} 
(\ref{DRS''''}) for 
the fermionic fields 
is the {\em bosonic} equation. Indeed, by substitution of the gauge field 
equations (\ref{fir'}) and (\ref{torcon}) and
a Fierz reordering,  the integrability condition 
\be 
D\left(\gamma\wedge D\Psi-{1\over 2}\,\Theta\wedge\Psi\right) 
=D\left(\gamma\wedge D\Psi\right)=0 
\label{gauge'}\,  
\ee 
is satisfied without having used explicitly the RS field equation  
(\ref{RS-EC'}), 
but implicitly employed it in the energy--momentum current 
because of $L_{\rm RS}\cong 0$. 

\subsection{SUSY transformation}
The integrability condition 
(\ref{gauge'}) points to a {\em gauge invariance} of simple 
supergravity. The  
invariance of $L_{\rm Sugra}$ under such local 
{\em supersymmetric transformations} (SUSY)  \cite{DZsugra}
becomes now rather natural by using the Clifford algebra--valued  
coframe and connection. They read
in our notation: 
\ba 
\delta\overline{\Psi}&=&2\overline{D\epsilon}\,,\nonumber \\ 
\delta\gamma &=& i\ell^{2}\overline{\epsilon}\gamma^\alpha\Psi\gamma_\alpha\,,
\nonumber \\ 
\delta \Gamma^{\star \alpha}&=&
i\ell^{2}\overline{\epsilon}\gamma_5\gamma^\alpha D\Psi 
\label{varpot}\, , 
\ea 
where  
$\delta \Gamma^{\star \alpha}:={1\over 2}\,\delta\Gamma^{\beta\gamma}\wedge
\eta_{\alpha\beta\gamma}$ corresponds to the Lie dual of the 
variation of the connection, and
$\epsilon^\alpha(x)$ is an anticommuting spinor--valued  
zero--form (Majorana field). 

Quite generally, we find for the variation of the supergravity  
Lagrangian 
\ba 
\delta L_{_{\rm Sugra}}&=&\delta\vartheta^\alpha\wedge\left[{1\over 2\ell^{2}} 
\eta_{\alpha\beta\gamma}\wedge R^{\beta\gamma}-{1\over 2}\, 
(\overline{\Psi}\gamma_5\gamma_\alpha\wedge D\Psi+\overline{D\Psi} 
\wedge\gamma_5\gamma_\alpha\Psi)\right]\nonumber\\ 
& &+{1\over 2}\,\eta_{\alpha\beta\gamma}\wedge\left(-{1\over  
\ell^{2}}\, 
T^\gamma+{i\over 2}\,\overline{\Psi}\wedge 
\gamma^\gamma\Psi\right)\wedge\delta\Gamma^{\alpha\beta}\nonumber \\ 
& &+\delta\overline{\Psi}\wedge\gamma_5\left(i\gamma\wedge D\Psi- 
{i\over 2}\,\Theta\wedge\Psi\right)\,.
\label{varlagr}
\ea 
Substituting (\ref{varpot}), this becomes 
\ba 
\delta L_{_{\rm Sugra}}&=&i\ell^{2}\overline{\epsilon}\gamma^\alpha\Psi\wedge 
\left[{1\over 2\ell^{2}} 
\eta_{\alpha\beta\gamma}\wedge R^{\beta\gamma}-{1\over 2}\, 
(\overline{\Psi}\gamma_5\gamma_\alpha\wedge D\Psi+\overline{D\Psi} 
\wedge\gamma_5\gamma_\alpha\Psi)\right]\nonumber\\ 
&+& i\overline{\epsilon}\gamma_5\gamma_\alpha D\Psi\wedge 
\left(T^\alpha-{i\ell^{2}\over 2}\,\overline{\Psi}\wedge 
\gamma^\alpha\Psi\right)\nonumber\\
&+&2\overline{D\epsilon}\wedge\gamma_5 
\left(i\gamma\wedge D\Psi- {i\over 2}\,\Theta\wedge\Psi\right) 
\label{varlagr'}\, . 
\ea 
By means of (\ref{DRS''''}) and a Fierz reordering, this turns out to  
be a boundary term, which vanishes `on shell':  
\be 
\delta L_{_{\rm Sugra}} =2d\left\{\overline{\epsilon}\wedge\gamma_5 
\left(i\gamma\wedge D\Psi- {i\over 2}\,\Theta\wedge\Psi\right)\right\} \cong 0 
\label{vargauge}\,. 
\ee 
Therefore,  we have explicitly shown that simple supergravity is {\em locally}  
supersymmetric; the vanishing of the covariant derivative of (\ref{RS-EC'}), 
provided that (\ref{fir'}) and (\ref{sec2}) are fulfilled, accounts for it. 
Simple supergravity leads to a generalization of the Dirac equation  
for the 
{\em massless gravitino}, which has the virtue of being free of the 
Velo--Zwanziger causality problems of other higher spin equations. 

Now according to our results concerning the chiral Clifford 
algebra--valued formulation there is in the Lagrangian density 
$\stackrel{(\pm)}{V}_{\rm EC}$ a term $\sim Tr (\Theta\wedge\Theta)$, 
quadratic in the torsion. 
This term vanishes for supergravity due to the following Fierz rearrangement
similar to (\ref{torcon}): 
\begin{eqnarray} 
Tr (\Theta\wedge\Theta) 
&=& -{\ell^{4}\over 4}Tr\left({\overline{\Psi}}\wedge  
\gamma^{\alpha}\Psi 
\wedge {\overline{\Psi}}\wedge \gamma_{\alpha}\Psi \right)\nonumber \\
&=&-{\ell^{4}\over 4}Tr\left({\overline{\Psi}}\wedge {\overline{\Psi}}\wedge 
\gamma^{\alpha}\Psi \wedge \gamma_{\alpha}\Psi \right)\nonumber \\
&=& 0\, . 
\end{eqnarray} 
\section{Chiral supergravity induced via a translational Chern--Simons term}

In order to give the supergravity Lagrangian (\ref{eq:sugra}) 
its chiral form, an analysis similar to the one given in the previous 
section should be performed. Jacobson \cite{Jacob88} advanced to  
some extent in this direction and there exists also a twistor formulation \cite{Ge98};
here we focus on  the more fundamental  
point of view of the generating function \cite{ma96}. 
To this end the Rarita--Schwinger spinor one--form is decomposed 
\cite{claus,claus1} into left and right--handed pieces 
$\Psi=\Psi_{\rm L} + \Psi_{\rm R}$, 
where 
\be 
\Psi_{\rm L}:= {{1-\gamma_5}\over 2}\, \Psi=P_{-}\Psi\, ,\qquad  
\Psi_{\rm R}:= 
{{1+\gamma_5}\over 2}\, \Psi=P_{+}\Psi\, , 
\ee 
with $\overline{P_{\mp}\Psi}={\overline \Psi}P_{\pm}$, 
$D_{\pm}\Psi_{\rm L,R}:=d\Psi_{\rm L,R}+\Gamma_{\pm}\wedge \Psi_{\rm L,R}$,
and 
$\overline{D_{\pm}\Psi_{\rm L,R}}:=d\overline{\Psi_{\rm L,R}}+ 
\overline{\Psi_{\rm L,R}}\wedge\Gamma_{\pm}$. 

In our elegant ``{\em Clifform}" approach, we note that `on shell',  
i.e. after using the Cartan relation (\ref{sec1}), the 
translational Chern--Simons term (\ref{CTT}) is given by 
\begin{equation} 
C_{\rm TT}\cong {i\over 4}\, 
{\overline{\Psi}}\wedge  {\gamma}\wedge{\Psi}= {1\over 4}\,J_5\, , 
\end{equation} 
which is proportional to the {\em axial current} 
$J_5:=i{\overline{\Psi}}\wedge  {\gamma}\wedge{\Psi}$ of the  
Rarita--Schwinger field. 

Similarly as in the Dirac case, the chiral version of the  
Rarita--Schwinger Lagrangian is obtained \cite{MMM96,mie86} by adding 
the boundary term 
$dC_{\rm TT}$ with the imaginary unit:
\ba 
L_{\rm RS\pm}&:=& {1\over 2} L_{\rm RS} \mp idC_{\rm TT}\nonumber \\ 
&=& \pm  
\frac{1}{2}\left(\overline{D\Psi}P_{\pm}\wedge\gamma\wedge\Psi+ 
\overline{\Psi}\wedge P_{\mp}\gamma\wedge D\Psi\right)\, . 
\ea 
Note that in the  `on shell' expression of the boundary term the torsion
$\Theta=D\gamma$ drops out due to (\ref{torcon}). 
Acting with the chirality projector on the spinor one--forms as well as on the
connection, the chiral Rarita--Schwinger Lagrangian reads explicitly 
\be 
 L_{{\rm RS}\pm} = \pm {1\over 2}\left\{\overline{\Psi_{\rm R,L}}\wedge
 \gamma\wedge D_\mp\psi_{\rm R,L} + 
 \overline{D_\pm\Psi_{\rm L,R}} 
  \wedge\gamma\wedge\Psi_{\rm L,R}\right\}\,, 
\label{chirs} 
\ee 
where the positive (negative) sign goes once again with the first  
(second) of the indices appearing in the vector--spinor $\Psi_{L,R}$. 
According to this result, the chiral and standard actions for supergravity 
differ by an imaginary boundary term, once the equation 
of motion for the connection obtained from the chiral one 
is used. 
In full, the {\em chiral} supergravity theory can be related to simple   
supergravity through the identity 
\be 
L^{\rm Chiral}_{\rm Sugra}:={1\over 2}\stackrel{(\pm)}{V}_{\rm EC}+ 
L_{\rm RS\pm} = {1\over 2}\left(L_{\rm Sugra} \mp i\,dC_{\rm TT}\right)\,. 
\label{chsugra}
\ee 
More precisely, the imaginary part of the chiral 
supergravity Lagrangian (\ref{chsugra}) is the boundary term 
\begin{equation} 
 L^{\rm Chiral}_{\rm Sugra} - {\overline L^{\rm Chiral}_{\rm Sugra}} 
=\mp i dC_{\rm TT}\cong i\left[\pm {i\over 4} d\left({\overline{\Psi}}\wedge 
{\gamma} \wedge{\Psi}\right)\right]\, , 
\end{equation} 
whereas the real part is the standard supergravity Lagrangian  
\cite{Jacob88,ma96}. 

\subsection{SUSY invariance of the translational boundary term}
Let us finally  check, if the additional boundary term $dC_{\rm TT}$ preserves local 
supersymmetry `on shell': For
the variation of the translational Chern--Simons term $C_{\rm TT}$ we can use 
the identity $\delta \left(\vartheta^\alpha \wedge T_\alpha\right) = 
\delta \vartheta^\alpha
\wedge 2T_\alpha - d\left(\vartheta^\alpha \wedge \delta \vartheta_\alpha
\right)$.  
Under the usual assumption that $[\delta,d]=0$, the variation of 
the corresponding boundary term yields
\be
\delta dC_{\rm TT} = \frac{1}{2\ell^2}\left[ d \left(\delta\vartheta^\alpha \wedge 
2T_\alpha\right) - dd \left(\vartheta^\alpha \wedge \delta\vartheta_\alpha
\right)\right] = \frac{1}{\ell^2} d\left(\delta \vartheta^\alpha 
\wedge T_\alpha\right)
\label{egg}\, .
\ee
Thus, for the SUSY transformations (\ref{varpot}) we find 
\ba
\delta dC_{\rm TT} &=& i d\left(\overline{\epsilon}\gamma^\alpha \Psi \wedge 
T_\alpha\right) \nonumber \\
&\cong& \frac{\ell^2}{2}d\left(\overline{\epsilon}\gamma^\alpha \Psi \wedge 
\overline{\Psi}\wedge \gamma_\alpha \Psi\right)  \nonumber \\
&=& 0
\label{yuyu}\, . 
\ea
In the last step we used the Cartan relation (\ref{sec2}) and a Fierz
rearrangement of two $\gamma^\alpha \Psi$ one--forms.
Consequently, the translational Chern--Simons term does {\em not} spoil the
supersymmetry for our {\em chiral} formulation of supergravity. 
Hence, the two theories are dynamically equivalent, but have  
canonically transformed variables in the Hamiltonian formulation. 

Moreover, in the chirally projected form the SUSY transformations read
\ba 
\delta\overline{\Psi_{\rm L,R}}&=&2\overline{D_\pm \epsilon_\pm}\,,\nonumber \\ 
\delta\gamma_\pm &=& i\ell^{2}\overline{\epsilon}_\pm\gamma^\alpha\Psi_{\rm L,R}\gamma_\alpha\,,
\nonumber \\ 
\delta \Gamma^{\star \alpha}_\pm&=&
i\ell^{2}\overline{\epsilon_\pm}\gamma_5\gamma^\alpha D_\pm\Psi_{\rm L,R} 
\label{chivarpot}\, . 
\ea 
For an extended model of chiral supergravity with {\em complex} tetrads, 
left- and right-handed SUSY transformations with 
$\delta_{\rm L,R}\Psi_{\rm R,L}=0$ 
have tentatively adopted \cite{Tsu1}, 
in order to avoid inconsistencies. However, by adopting 
our Chern--Simons type 
boundary term \cite{mie86,MMM96} for the imaginary part (\ref{chsugra})
of chiral supergravity the consistency problem can be solved also for 
(${\cal N}$=2), see \cite{TS98}.

\section{Chiral anomaly} 
{}From the Dirac equation and its conjugate one can readily deduce for the  
{\em axial current} $j_5 := \overline{\psi}\gamma_5 \, ^*\gamma\psi$ that
the well--known ``classical axial anomaly"
\cite{kak} 
\be dj_5 =2imP  =2im \overline{\psi}\gamma_5\psi
\ee
for {\em massive} Dirac fields $\psi$ 
holds also in a Riemann--Cartan background spacetime. 
If we restore chiral symmetry in the limit   
$m\rightarrow 0$, this leads to classical conservation law of the 
axial current for massless Weyl spinors, or since $dj :=
d \overline{\psi}\,^*\gamma\psi=0$  , equivalently, for the 
{\em chiral currrent} $j_\pm :=  
\overline{\psi}(1 \pm\gamma_5)\,^*\gamma\psi/2=\overline{\psi}P_\pm\,^*\gamma
\psi$.   

However, within the dynamical framework of the Einstein--Cartan--Dirac theory 
we  also have
\be 
dC_{\rm TT}\simeq (1/4) dj_5 \rightarrow 0 
\ee
`on shell'. This is consistent with the fact that a 
Weyl spinor does not couple to torsion, because  the remaining axial torsion 
$A:=\,^*(\vartheta_\alpha\wedge T^\alpha)$ becomes a {\em lightlike} 
covector, i.e.
$A_\alpha A^\alpha\eta =A\wedge\,^* A \simeq (\ell^4/4)\,^*j_5\wedge j_5 =0$.   

It is therefore worth mentioning at this point that the  Ashetekar type
dynamical equivalence of the chiral formulation 
holds at the classical level because 
$dC_{\rm TT}\cong 0$ `on shell', at least for massless spinors. 

When quantum field theory is involved, other boundary terms 
may arise from  the {\em chiral anomalies} due to the  
non--conservation of the axial current,  cf. \cite{zum,Hir}. In our case, 
its calculation is much facilitated by 
regarding one term in the decomposed Dirac Lagrangian 
as an {\em external} axial covector $A$ (without 
``internal" indices)  coupled to
the axial current $j_5$ of the Dirac field in an {\em initially flat} 
spacetime. Then we can apply
the result (11--225) of Itzykson and Zuber \cite{IZ} for the axial anomaly. 
Accordingly, we find that only the term 
$dA\wedge dA$ arises in the  
axial anomaly \cite{MK}, but {\em not} the Nieh--Yan type term 
$d\,^*A\sim dC_{\rm TT}$ as was 
recently claimed \cite{ChandiaZ}.
After switching on the curved spacetime of Riemannian geometry, we finally  
obtain for the axial anomaly
\be
<dj_5>= 2im <P> + {1\over 24\pi^2}\left[Tr\left(\Omega^{\{\}}\wedge  
\Omega^{\{\}}\right)  -{1\over 4} dA\wedge dA\right]\,.
\ee
This result, which can easily be transfered to the chiral current $j_\pm$, 
is based on diagrammatic techniques and 
the Pauli--Villars regularization scheme. 
It  deviates sharply 
from  the heat kernel method \cite{Yuri1,Yuri2,Yuri3} which seems to lead to  partially 
{\em divergent} terms, and thus cannot be applied to our problem.

In the limit $m\rightarrow 0$, one can easily read off the corresponding 
chiral anomaly.   
For gravitational Rarita--Schwinger fields $\Psi$ within 
supergravity, the anomaly \cite{Gr78,CD78} for the 
{\em axial current} $J_5 := i\overline{\Psi}\wedge \gamma\wedge\Psi$ 
is $-21\times$ the anomaly for 
Dirac fields, whereas for the corresponding supersymmetric Yang--Mills 
anomaly one finds $3\times$ the Dirac result.

$$\vbox{\offinterlineskip
\hrule
\halign{&\vrule#&\strut\quad\hfil#\quad\hfil\cr
height2pt&\omit&&\omit&&\omit&\cr
& {\bf Spin }  &&{\bf Gravitational} && {\bf YM anomaly}&\cr
height2pt&\omit&&\omit&&\omit&\cr
\noalign{\hrule}
height2pt&\omit&&\omit&\cr
& 1/2&& 1 && 1  &\cr 
& 3/2&& $-21$ && 3  &\cr  
height2pt&\omit&&\omit&&\omit&\cr}
\hrule}$$

Depending on the asymptotic helicity states, there occur  
contributions of topological origin of the Riemannian 
Pontrjagin or Euler type, 
respectively. Interesting enough, there is a Pontrjagin type contribution   
$dA\wedge dA$ from axial torsion in Riemann--Cartan spacetime.  
The role of spinors for the index theorem and in the $4D$ Donaldson invariants
via Seiberg--Witten equation has recently been reviewed by Atiyah
\cite{At96}.
\section{Reality conditions}

The canonical variables introduced by Ashtekar cast the constraints of general 
relativity into polynomial form. This is a major achievement since the general
relativity constraints when written in the usual ADM variables are highly
non--polynomial and not even analytic in these variables and this has been
one of the major difficulties in quantizing canonical gravity. The simplicity of
the Ashtekar formulation gets afflicted by the complex character of his
variables, the use of reality conditions is compulsory. The pursuit of the
Ashtekar program was formulated as an alternative route for classical general 
relativity and its canonical quantization. In the classical theory,
general relativity is considered to be embedded in a larger complex theory;
the restriction to the usual Einstein theory is made imposing by hand some
reality conditions. In quantum theory, one first ignores the reality conditions, 
solves the quantum constraints of the complex theory and  the 
reality conditions finally surface as constraints upon the admissible inner products by
requiring that the real classical observables become self--adjoint operators.
The use of the reality conditions as a way of selecting an inner product
has worked well in some models, but the same has not happened for the full
gravity or supergravity sofar. 

\subsection{One--dimensional harmonic oscillator analog}

To illustrate the procedure of the last sections 
on a most elementary level, we take the one--dimensional 
harmonic oscillator of one Hertz 
as a toy model and show how the complex formulation of supergravity
should work a la Dirac and why the Ashtekar formulation does not allow in a 
straightforward way the definition of an inner product needed to enforce
 the reality conditions.

\subsubsection{Dirac factorization method}
Let us recapitulate the factorization scheme introduced by 
Dirac  and Bargmann for obtaining the eigenstates of the 
one--dimensional harmonic oscillator. Classically, its Lagrangian  reads
 \begin{equation}  {\cal L}={1\over 2}{\buildrel \bullet\over q}^2 -U(q)= 
{1\over 2}({\buildrel \bullet\over q}^2 -q^2)
\, ,\end{equation}  
where  $\bullet:=\partial/\partial t$  denotes the time derivative.
In terms of the canonical momentum $p:=
{\partial{\cal L}/{\partial{\buildrel \bullet\over q}}} =
{\buildrel \bullet\over q}$  the 
Hamiltonian takes the form
\be
{\cal H}=p{\buildrel \bullet\over q} -{\cal L}= 
{1\over 2}(p^2 +q^2)\, . \label{Ham}
\ee

In quantum mechanics (QM), the corresponding {\em annihilation and creation} 
operators $a$ and $a^\dagger$  satisfying $[a, a^\dagger] =1$ 
are defined as  follows:
\begin{equation}
a := \frac{1}{\sqrt{2}}\left(q + i\, p \right)\, ,
\quad
a^\dagger :=\frac{1}{\sqrt{2}}\left( q - i\,p  \right)\, . 
\end{equation} 
Here $q$ is the generalized coordinate operator 
and $p = \left( a - a^\dagger \right)/2i$ its canonical conjugate
momentum. The dagger is used to denote the adjoint.

The Hamiltonian (\ref{Ham}) converts into the operator
\begin{equation}
{\cal H} = a\, a^\dagger - \frac{1}{2}= a^\dagger\, a + \frac{1}{2}
= \frac{1}{2} \left(a\, a^\dagger + a^\dagger\, a\right) 
\end{equation}
and allows to write the 
 Schr\"odinger equation ${\cal H} \psi_E =E \psi_E$  in three diffrent but
completely equivalent ways.

As it is well known, the operator $a$ {\em annihilates} the
ground state $\psi_0=c_0\exp(-q^2/2)$ with energy $E_0=1/2$ of the 
harmonic oscillator, i.e. $a\,\vert \psi_0 \rangle  = 0 $. 

The remaining states can be constructed by successive application
of the {\em creation operator} $a^\dagger$ to $\psi_0$, 
the energy increasing at each step by one, i.e.
\begin{equation}
\psi_n = (n!)^{-1/2}\left( a^\dagger \right)^n \psi_0
= A_n \, H_n \left(q\right) \exp\left(- q^2/2\right)\, .
\end{equation}
Here $c_0$ and $A_n$ are normalization constants, $H_n(q)$ the Hermite
polynomials, and $E_n = n + \frac{1}{2}$ 
 the eigenvalues of the energy.
 
In this case there exist a well--defined inner product 
\begin{equation}
\langle \psi_m \vert \psi_n \rangle  = (n! m!)^{-1/2}
 \langle \psi_0\vert a^m a^{\dagger n}\vert
\psi_0\rangle =  (n! m!)^{-1/2} n!\, \delta_{mn}
\langle \psi_0\vert\psi_0\rangle = \delta_{mn}\, ,
\end{equation}
which allows real expectation values of  self--adjoint
operators like the kinetic energy operator:
\begin{equation}
\langle \psi_n \vert \frac{p^2}{2}\vert \psi_n \rangle =
-\frac{1}{4}\langle \psi_n \vert \left( a - a^\dagger \right)^2
\vert \psi_n\rangle
=\frac{1}{4}\langle \psi_n \vert a\, a^\dagger + a^\dagger\, a\vert 
\psi_n\rangle
= \frac{1}{2}\left( n + 
\frac{1}{2} \right)= 
\frac{1}{2}\,E_n\, .
\end{equation}
It is important to stress the fact that in the Dirac's factorization method
both variables, i.e. the generalized coordinate ant its canonical conjugated
momentum, are {\em complex}. Although this is a complex formulation of
the harmonic oscillator, it leads to Hermitian operators having
{\em real} eigenvalues. In contrast, in  the Ashtekar formulation of
 the next section, only the canonical momenta 
become complex while the generalized coordinates, the triad, remain real,
leading to non--Hermitian operators. Moreover, until now there does not exist a well
defined inner product in order to obtain real eigenvalues.

\subsection{Hybrid representation}

In a `nutshell', the Ashtekar approach to quantum gravity proceeds from 
a {\em canonical transformation}
$(q,p)\rightarrow (\widetilde q, \widetilde p)$  induced by 
${\cal C}=q^2/2$ as  
{\em generating function}. On the Lagrangian level of our toy model
 this is equivalently 
achieved by adding the boundary term  
$i{\buildrel \bullet\over {\cal C}}$, resulting  in the 
complex Lagrangian
\begin{equation} {\buildrel \pm \over {\cal L}}=
{\cal L} \pm i{\buildrel \bullet\over {\cal C}}= 
{\cal L}\pm i{\buildrel \bullet\over q}\, q \,.  \label{comLag}
\end{equation}  

The corresponding {\em complex momenta} are 
${\buildrel \pm \over p}:= 
{\buildrel \bullet\over q} \pm i
\partial{\buildrel \bullet\over {\cal C}}/
{\partial{\buildrel \bullet\over q}} =
{\buildrel \bullet\over q} \pm i
\partial {\cal C}/
{\partial q} 
=p \pm iq\quad [=\pm i\sqrt{2}a^{\>(\dagger )}]$.
This resembles the 
Bargmann representation of the harmonic oscillator, but only partially, 
because merely  one choice of the complex momenta, for instance $p-iq$,
 but not its conjugate will be employed
 in the Ashtekar formulation.  In this {\em `hybrid'} representation,
the   Hamiltonian
${\buildrel \pm \over {\cal H}}:={\buildrel \pm \over p}\,
{\buildrel \bullet\over q} - {\buildrel \pm\over {\cal L}}= 
{1\over 2}{\buildrel \pm \over p}\,{\buildrel \pm \over p} -
i {\buildrel \pm \over p}\,q = {\cal H}$ 
turns out to be  holomorphic in ${\buildrel \pm \over p}$.

In QM, where $q$ and $p$ become operators obeying $[q,p]=i$, 
the canonical transformation is achieved by 
the {\em similarity transformation}
\begin{equation} \cases{\widetilde q = N q N^{-1} =q & \cr 
\widetilde p = N p N^{-1} =p \pm i
\partial{\buildrel\bullet\over{\cal C}}/{\partial{\buildrel 
\bullet\over q}} = {\buildrel \pm \over p}& \cr} \Rightarrow 
N= \underbrace{e^{\pm {\cal C}}}_{\rm non-unitary}\end{equation}  
The construction of the {\em non--unitary} operator $N$  proceeds via
$e^{\cal C} pe^{-\cal C} = p + [{\cal C}, p] + 
{1\over 2!} [{\cal C}, [{\cal C}, p]] +
{1\over 3!}  [{\cal C}, [{\cal C}, [{\cal C}, p]]] + \cdots 
=p \pm iq$.
Since $[q^2, p] =q [q,p] + [q,p] q = 2i q$, we necessarily  recover  
${\cal C}= q^2/2$.  Observe that in the Schr\"odinger representation the addition of the 
`complexifier' $i{\buildrel \bullet\over {\cal C}}$ in the Lagrangian 
(\ref{comLag})  induces a  {\em renormalization} of the wave function
\begin{equation} \psi =N\widetilde\psi = \exp\left (\pm q^2\right)
\widetilde\psi\, ,\quad \widetilde\psi_{n}=A_n\, 
H_n\left(q\right)\, .\end{equation}  

In the $\widetilde p$ representation the Schr\"odinger equation becomes simplified to 
a first--order differential equation, but at the price of   
modifying the inner product by a {\em nonlocal} measure \cite{FK90}.
With a
{\em Wick rotation} in phase space one can go back to the usual 
creation and annihilation operators $a$ and 
$a^{\dagger }$ and retain an inner product for implementing the reality conditions, 
 see Ref. \cite{thie}.

 The Hamiltonian formulation of gravity 
 can be easily implemented in our scheme following the usual Yang--Mills or Poincar\'e gauge description, see 
Refs. \cite{[13],ma,BM}. 
In the Ashtekar formulation  with complex variables, the  
tangential part of the basis one--forms, i.e. the
{\em  ``triad densities"}, and the tangential part of the self-- or 
antiself dual connection connection will 
become the generalized coordinates  
and momenta of the bosonic sector. In supergravity,
the {\em tangential} Rarita--Schwinger field will be   
on par with the triads, see Ref. \cite{MMS98}   for details.
In the context 
of EC theory, Maluf \cite{Ma92} 
has argued that the reality condition leads to the vanishing of the 
spacelike torsion; but this would 
clash with the algebraic torsion relation (\ref{sec2}) in supergravity.

There have been many attempts to solve the reality conditions issue,
namely by considering ``generalized Wick transform'' \cite{thie,wick},
by imposing  the reality conditions, not by 
hand, but via Dirac constraints \cite{hamt}, or by using a modified form 
of the self--dual action that leads to the $SO(3)$ ADM formalism \cite{barb95}
without the appearence of difficult second--order constraints. However,
there is a price to be paid, the Hamiltonian constraint is no longer a
simple quadratic expression in both the densitized triad and the
Ashtekar connection. This fact makes it more difficult to discuss all those
issues that depend critically on having the theory formulated in terms of 
simple constraints; in particular, solving the constraints will be harder now.
Therefore, the structure of the Hamiltonian constraint is at least as
complicated as the one of the familiar ADM constraint \cite{henn89}.

One interesting point of discussion, suggested by the lack of success of all
the attempts to retain   Lorentzian general relativity by introducing simple
modifications in the know actions, has to do with the obvious asymmetry 
between the formulations of gravity or supergravity in a real Ashtekar 
phase space for Lorentzian and Euclidean signatures. In fact it all boils
down to the relative signs between the potential and kinetic terms in the
scalar constraint. Consequently, the origin of the signature at
the Lagrangian level is also rather obscure. The marked asymmetry 
between the real Hamiltonian formulation for different spacetime
signatures strongly suggests that it would differ very much from the usual
self--dual action. This could have intriguing consequences in a
perturvative setting because the ultraviolet behavior of the Euclidean
and the Lorentzian theories could be very different. It is worthwhile to
stress  this point for  the Hilbert--Einstein action and the so--called
higher derivative theories, that differ in some terms quadratic in the 
curvatures. The first one is
nonrenormalizable, whereas the second one is perturbatively 
renormalizable but nonunitary \cite{St77}.   

In full, the reality conditions issue deserves further investigation.

\section{Discussion} 

As we have shown, there exists a natural way to translate (${\cal N}$=1) supergravity  
theory into chiral form, by adding certain Chern--Simons boundary  
terms to the original action. This prescription
works elegantly not only for the case of chiral simple (${\cal N}$=1)  
supergravity, but also for the Einstein--Cartan--Dirac theory,
see  \cite{MMM96} for details.

For the bosonic sector, the boundary term (\ref{CTT}) coming from the 
translational Chern--Simons term (\ref{CTT}) has the usual  
geometric form.
In the case of the gravitino spinor field such boundary term is  
proportional to the axial current and arises naturally by using the algebraic 
Cartan relation as the second field equation.
Since standard Rarita--Schwinger Lagrangian is zero `on shell'
it is not surprising that the same happens in their chiral form.  
Consistenly, the correspondent boundary terms relating both forms of the above
Lagrangian semi--classically turn out to be zero `on--shell' again, i.e. 
$dC_{\rm TT}\cong 0$,   at least 
for massless spinor field.  Consequently, the
addition of a suitable Chern--Simons boundary terms seems to be a  
unifying principle in order to obtain the chiral formulation of fermions  
coupled to gravity.

Spinor formulations of general relativity have become very  
popular again. For instance, Nester and Tung \cite{Ne95} invented an  
interesting quadratic spinor Lagrangian for general relativity. However, the  
variational principle seems to be incomplete, since the normalization  
$\overline{\Psi} \Psi=1$ of the Rarita--Schwinger type one--form $\Psi$ is not
consistenly included via a Lagrange multiplier two--form. 

In  previous papers of Tsuda et al. \cite{Tsu,Tsu1} on (${\cal N}$=1) and 
(${\cal N}$=2) supergravity, boundary terms are  
consistenly neglected thus completely missing the crucial role of the  
translational Chern--Simons term as the 
generating function for the chiral variables. In a  
recent work of Tung and Jacobson \cite{TJ}, our earlier differential form 
approach including a translational Chern--Simons term has been  
partially recovered in spinor notation without, 
however, using the unifying notion of Clifford algebras.  
For massive fermions in a gravitational field with angular momentum, there 
occur chirality transitions, see \cite{CM94}, which in extreme astrophysical 
situations could give way to {\em sterile} particles. A recent paper 
\cite{EK96} seems to indicate that simple supergravity may not even be 
one--loop finite in the presence of boundaries.

 \section*{Acknowledgments} 
We would like to thank Hugo Morales--T\'ecotl,  Yuri Obukhov  as well as the referee 
for useful hints and comments.
This work was partially supported by  CONACyT, grant No. 28339E, and the 
joint German--Mexican project DLR--Conacyt
 E130--2924 and MXI 009/98 INF. 
E.W.M. acknowledges the honorary membership at IRB, Castle Prince Pignatelli 
I--86075 Monteroduni (IS), Molise, and 
thanks Noelia M\'endez C\'ordova for encouragement.

\end{document}